# From Brain to Motion: Harnessing Higher-Derivative Mechanics for Neural Control

O. White[1], F. Buisseret[2,3], F. Dierick[2,4,5], and N. Boulanger[6]

1. Université Bourgogne Europe, INSERM-U1093 Cognition, Action, and Sensorimotor Plasticity, Campus Universitaire, BP 27877, 21078 Dijon, France;
2. CeREF-Santé, Chaussée de Binche 159, 7000 Mons, Belgium;
3. Service de Physique Nucléaire et Subnucléaire, Université de Mons, UMONS Research Institute for Complex Systems, 20 Place du Parc, 7000 Mons, Belgium;
4. RehaLAB, Centre National de Rééducation Fonctionnelle et de Réadaptation—Rehazenter, Rue André Vésale 1, 2674 Luxembourg, Luxembourg ;
5. Faculté des Sciences de la Motricité, UCLouvain, 1-2 Place Pierre de Coubertin, 1348 Louvain-la-Neuve, Belgium;
6. Service de Physique de l'Univers, Champs et Gravitation, Université de Mons, UMONS Research Institute for Complex Systems, Place du Parc 20, 7000 Mons, Belgium.

Optimal Feedback Control (OFC) provides a theoretical framework for goal-directed movements, where the nervous system adjusts actions based on sensory feedback [1,2]. In OFC, the central nervous system (CNS) not only reacts to stimuli but proactively predicts and adjusts motor commands, minimizing errors and (often energetic) costs through internal models. OFC theory can be extended beyond motor control to encompass perception and learning [3]. This theory assumes that there exists a cost function that is optimized throughout one's movement. It is natural to assume that mechanical quantities should be involved in cost functions. This does not imply that the mechanical principles that govern human voluntary movements are necessarily Newtonian. Indeed, the undisputed efficiency of Newtonian mechanics to model and predict the motion of non-living systems does not guarantee its relevance to model human behavior. We propose that **integrating principles from Lagrangian and Hamiltonian higher-derivative mechanics, i.e. dynamical models that go beyond Newtonian mechanics, provides a more natural framework to study the constraints hidden in human voluntary movement within OFC theory**. Such an integration is displayed in Figure 1 and will be extensively discussed hereafter. The outcome of our comment will be a refined framework for OFC that considers recent analyses based on Lagrangian or Hamiltonian mechanics, and that unifies them in a consistent way.

Newtonian mechanics obviously helps describe the immediate causal relationships from initial conditions to generation of a trajectory through the action of external forces – recall that according to standard terminology, internal forces are such that their sum always vanishes. However, Newtonian mechanics is not directly suited to address two primary challenges in understanding motor control:

modeling the redundancy inherent in the musculoskeletal system and the ability to perform voluntary movements against external loads, such as those induced by gravity. Redundancy refers to the multiple ways in which the same movement or task can be performed due to the numerous degrees of freedom available in our joints and muscles, and the question of how an individual is able to select a particular movement. Lagrangian mechanics is a priori well suited to model redundancy: It asserts that, out of all possible trajectories between two positions, $\vec{x}(t_i)$ and $\vec{x}(t_f)$, the actual path followed by the system under study is the one that minimizes an action functional $S[\vec{x}] = \int_{t_i}^{t_f} L(\vec{x}, \dot{\vec{x}}, \vec{x}^{(2)}, \ldots, \vec{x}^{(N)}) \, dt$, where $\dot{\vec{x}}$ and $\vec{x}^{(N)}$ denote the first and $N^{th}$ time-derivatives of the dynamical variables $\vec{x}$, respectively. The function $L$ is called the Lagrangian and the actual motion is given by Hamilton's variational principle, $\delta S = 0$, supplemented by a total of $2N$ conditions $\vec{x}^{(j)}(t_i) = 0 = \vec{x}^{(j)}(t_f)$, $j = 0, 1, \ldots, N-1$, at both initial and final times, which leads to the least-action trajectory [4]. **The action functional may therefore be a basic principle of movement planning**, i.e. it provides a principle by which the CNS can leverage multiple pathways to achieve the same movement goal and eventually chooses the best one, the one for which action is minimal. In that sense, the action functional $S[\vec{x}]$ may be thought of as a cost function in the OFC jargon. Newtonian mechanics has only $N = 1$, with $L$ being the difference between kinetic and potential energies. Higher-derivative systems are such that $N > 1$, and Lagrangians that are quadratic in $\vec{x}^{(N)}$ lead to equations of motion of order $2N$ with a solution requiring $2N$ initial conditions to be specified[5]. Note that both the non-local, Hamiltonian principle, and the local integration of an order-2N differential equation require the same number $2N$ of conditions, either $N$ conditions at both $t_i$ and $t_f$ in Hamilton's principle, or $2N$ initial (i.e., at $t_i$) conditions at the level of the equation of motion.

The case $N = 1$ is ruled out by human's ability to perform voluntary movements against given external forces, i.e. to choose accelerations that are not proportional to the total external force applied. The motor command and sensory feedback processed by the CNS constantly control muscle forces, hence the acceleration of a given joint or limb. The ability of the muscles to separately fine-tune initial position, velocity and acceleration has been shown by Ueyama and colleagues[6], leading to the conclusion that the minimal value of the integer $N$ should be 2. An $N = 2$ action, $S[\vec{x}] = \int_{t_i}^{t_f} L(\vec{x}, \dot{\vec{x}}, \vec{x}^{(2)}) \, dt$, is then higher-derivative and four initial conditions must be specified in the solution of the equations of motion: from $\vec{x}^{(0)}(t_i)$ to $\vec{x}^{(3)}(t_i)$. Interestingly, in the field of motor control, movements exhibiting two-thirds power-law, such as natural tracing or even writing, demand at least $N = 2$ to be produced [7,8]. Therefore, **we propose that the action principle governing movement planning has to be higher-derivative**.

The seemingly complex nature of selecting a trajectory by minimizing an action raises the question of how the brain manages it. The CNS adapts and optimizes motor control by learning from previous movements. This learning involves continuously updating internal models to improve predictions of motor command outcomes, based on feedback and on initial conditions. Hence, an action can serve as a reliable implementation of an internal model if an individual can "learn" it through repeated trials and observations. A four-layer neural network model with 500 hidden units is already able to learn a Lagrangian from the observation of 600,000 randomly generated trajectories [9]. Furthermore, a small neural network can replicate reaching movements in monkeys, suggesting that the CNS, with its vast complexity, can learn and optimize similar tasks [10]. These network architectures being considerably simpler than the human brain, it is reasonable to assume that several Lagrangians may be learned concurrently by one individual. This implies that motor strategies, such as non-trivial cost functions, may be encoded in neural circuits.

Once the action is selected, the corresponding Hamiltonian, *H*, follows. Let us clarify the relevance of Hamiltonian mechanics at this stage. In higher-derivative mechanics, the *N* positions ($\vec{Q}$) and *N* momenta ($\vec{P}$) degrees of freedom can be computed from *L*, see e.g. Boulanger and colleagues for formal developments[8]. The space spanned by the pairs of variables ($Q^b, P_b$) is called phase-space (*b* denotes the vector component), in which the full system trajectory forms a single curve whose properties can be geometrically studied. To initiate the planned movement after the choice of higher-derivative action is made and the optimal trajectory selected, initial conditions on all the *2N* higher-derivative degrees of freedom are designed, that will trigger the movement in compliance with Hamilton's equations in the higher-derivative phase space. This process is illustrated in Fig. 1 (Planning and Control Policy boxes). **Therefore, Hamilton's equations coupled with appropriate initial conditions determine the attractor to be followed for the motor control**.

Phase-space provides a detailed map of the energy landscape of trajectories, guiding the CNS in its selection of the most efficient movement patterns. In an OFC framework, the optimal state estimator uses forward model to convert motor commands into estimates of limb position. **We propose that one individual can "store" phase-space trajectories corresponding to a given dynamics and initial conditions, called attractors hereafter, and exploit them as forward models, i.e. predicted attractors**. It predicts sensory consequences of motor commands and compares them to actual feedback to calculate the error. This iterative process occurs in real-time, allowing the CNS to refine the trajectory dynamically by comparison between the planned trajectory and the actual one. In our framework, the comparison is made between the predicted and measured attractors, that may reveal potential discrepancies in every degree of freedom. This naturally addresses the responsibility assignment problem, as the approach helps identify which of the many degrees of freedom diverge from the target trajectory [11]. By updating initial conditions without changing the dynamics, i.e. by updating internal models through sensory feedback in an OFC language, the CNS optimizes short-term adaptation. This mechanism may be related to error-based learning.

Long-term adaptation can occur by comparison of parameters relevant at longer time scales. In the case of a rhythmic movement for example, attractors become closed loops in phase-space planes $(Q^b, P_b)$. The areas of these loops, called adiabatic invariants and denoted $I$, are known to be constant in Hamiltonian mechanics and even remain weakly fluctuating in presence of noise. They also remain roughly constant in the presence of radically new environments, such as rapid transitions between different gravitational fields [12]. In other words, $\dot{I} \approx 0$. This constraint may allow the CNS to refine its control strategies over longer timescales by minimizing errors related to the fluctuations of $I$, thereby improving performance over a longer time scale than the process based on the attractor. We think that such long-range adaptations may even result in a change in the action functional and chosen Hamiltonian, through use-dependent learning mechanism. As an illustration, Raffalt et al. (2020) have shown that the walk-run transition may be seen as the transition toward two attractors, the attractor for running becoming more stable (as assessed by Lyapunov exponent) at higher speed[13].

How does the brain implement higher-derivative mechanical principles? Decades of research have uncovered functional specialization within both cortical and subcortical regions. However, a significant gap persists between our ability to leverage the brain for application and our deeper understanding of how various brain areas contribute to these tasks [14]. Today, most prosthetic control systems, for instance, heavily rely on signals from the primary motor cortex (M1), largely because it is easily accessible. This reliance on M1 oversimplifies the complexity of brain functions. Traditional approaches, such as using brain imaging with region-of-interest analysis, often assume distinct functions for each area. In fact, many brain regions serve diverse and overlapping roles. For example, the insula is involved in a wide range of processes, despite these processes being very different in nature, such as empathy [15] and graviception [16]. In this view, examining a brain area in isolation overlooks the complex, context-dependent interactions that occur across the vast neural network. No single region operates independently; instead, functions are likely implemented by distributed networks, with each area contributing in a dynamically modulated manner. Task goals and environmental cues are integrated within the dorsolateral prefrontal cortex, forming initial movement intentions. The posterior parietal cortex combines multisensory inputs to construct a state estimate of the body (initial conditions) and refine the motor planning component. M1 generates motor commands, while the cerebellum and primary somatosensory cortex process sensory feedback to refine ongoing movements. Going beyond this highly simplified story will open new avenues in motor control research.

Our framework brings a new vantage point on motor control, regarding for example (1) Connection to neurological diseases and rehabilitation as e.g. gait in patients suffering from Parkinson's disease, for which fluctuations between consecutive strides have lost predictability[17]. This loss may be related to an inability to ensure the constancy of the action variable ($\dot{I} = 0$), eventually supporting the use of auditory rhythmic stimulations as an effective rehabilitation option, see Fig. 1; (2) Identification of new motor invariants from $N \geq 2$ action principles, as already done to recover two-thirds law[8]; (3) Use of adiabatic invariant theory as the most powerful way to model the response of a system to external perturbations. This will enhance the predictive power of OFC models for individuals moving in variable environments with factors such as vibrations, luminosity, acoustic environment or even gravity [12].

In the intricate dance of motor control, the brain is not just a reactive machine—it is an anticipatory maestro, fine-tuning every movement through layers of prediction, feedback, and adaptation. **By pairing higher-derivative mechanical concepts and components of optimal feedback control models, one is led to a comprehensive framework that can link the intricate neural computations at work in voluntary movement to mathematically elegant physical laws governing them.**

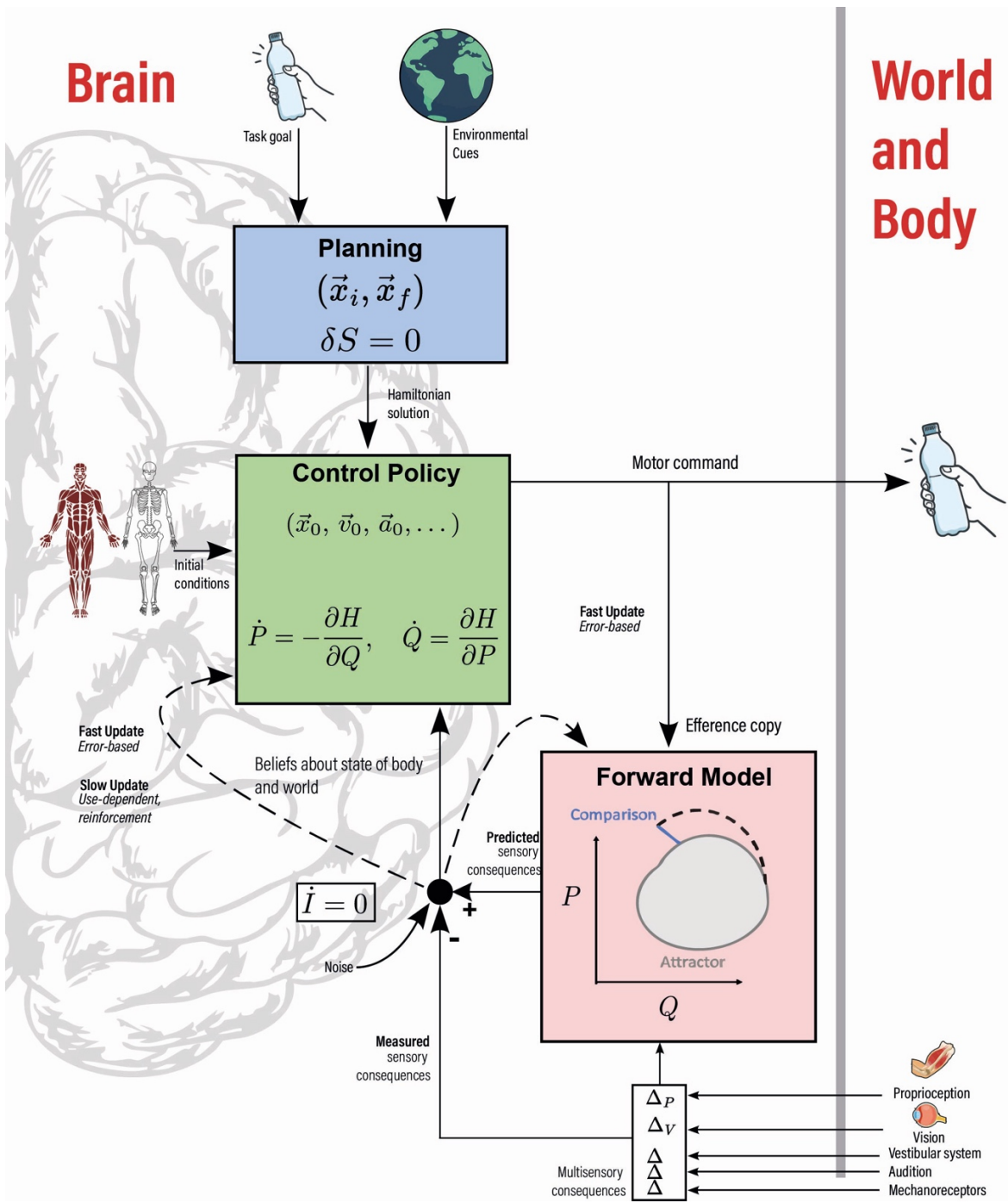

*Figure 1. Integration of higher-derivative mechanical concepts within components of an optimal feedback control model. It integrates task goals and environmental cues, such as gravity, into the planning process. A Hamiltonian equation of motion (green box), formulated through a higher-derivative least-action principle (blue box), informs the control policy. S and H represent the action and Hamiltonian respectively. Initial conditions and Hamilton's equations then guide movement execution. The optimal phase-space trajectory, or attractor, serves as a forward model (pink box), continuously compared to the actual trajectory. In rhythmic motion, the invariance of the adiabatic invariant (I) can act as a robust comparator. Movement updates occur rapidly through this comparison (error-based learning), while significant changes in task or environment may lead to longer-term adaptations in the variational principle (use-dependent learning). Illustrator: Robin Raedt.*